\documentclass[a4paper]{article}
\usepackage{blindtext, graphicx, fullpage}
\usepackage{subfigure}

\usepackage[justification=centering]{caption}

\usepackage{filecontents}
\usepackage[noadjust]{cite}
\usepackage{multirow}
\usepackage{enumerate}
\usepackage{ucs,enumerate}
\usepackage{multirow}
\usepackage{booktabs}
\usepackage{color,soul}
\usepackage{caption}
\usepackage{url}

\usepackage[cmex10]{amsmath}
\newcommand{\gaia}{\textsf{GAIA}\xspace}

\newcommand{\myparagraph}{\smallskip\noindent\textbf}

\begin{document}
%
\title{Scenarios for Educational and Game Activities using Internet of Things Data}

\author{Chrysanthi Tziortzioti \footnote{\textit{Hellenic Open University}, Patras, Greece, tziortzio@gmail.com},
Irene Mavrommati \footnote{\textit{Hellenic Open University}, Patras, Greece, mavrommati@eap.gr}, 
Georgios Mylonas \footnote{\textit{Computer Technology Institute \& Press}, Patras, Greece, mylonasg@cti.gr},\\
Andrea Vitaletti \footnote{\textit{Dep. of Computer, Control and Manag. Eng.}, \textit{Sapienza University of Rome}, Rome, Italy, vitaletti@diag.uniroma1.it}, Ioannis Chatzigiannakis \footnote{\textit{Dep. of Computer, Control and Manag. Eng.}, 
\textit{Sapienza University of Rome}, Rome, Italy, ichatz@diag.uniroma1.it}
}

\maketitle

\begin{abstract}
Raising awareness among young people and changing their behavior and habits concerning energy usage and the environment is key to achieving a sustainable planet. The goal to address the global climate problem requires informing the population on their roles in mitigation actions and adaptation of sustainable behaviors. Addressing climate change and achieve ambitious energy and climate targets requires a change in citizen behavior and consumption practices. 

IoT sensing and related scenario and practices, which address school children via discovery, gamification, and educational activities, are examined in this paper. Use of seawater sensors in STEM education, that has not previously been addressed, is included in these educational scenaria. 
\end{abstract}

\section{Introduction}
\label{sec:intro}
Wireless Sensor Networks have seen a tremendous development, leading to the realization of the Internet of Things (IoT). Today, there is a large variety of hardware and software to choose from that is easy to set up and use in an increasing number of real-world applications~\cite{Chatzigiannakis2011103,ewsn10}. One such application is education: the deployment of a variety of sensors (e.g., for monitoring electricity consumption, environmental conditions, be them indoor or outdoor, etc.) across school buildings or across different natural water reserves, during different seasons, can produce real-world data to be directly used in STEM educational activities.

This is a crucial period for the future of our planet as it becomes evident that human activities inflict irreversible damage on the environment and on critical resources. One approach for addressing the climate change problem is through the development and transfer of green technologies. Emphasis is given to Environmental awareness via STEM education. This is achieved by sensor experiments, utilizing off the shelf IoT sensors, and educational activities planned specifically for school children, as part of their Science class. Gamification elements are inseparable in such approaches: discovery and adventure are intemperate elements in child’s play that leads them through knowledge. Gamification via the internet and social activity mechanisms, on the other hand, is multiplying the impact of the children engagement.

This paper deals with energy and environmental awareness as a part of STEM school educational activities. This is handled in two ways: a) by addressing energy footprint and energy consumption, via individual and group class activities, by using IoT sensors and gamification elements using real sensor data from familiar environments and recording changes in behaviour that affect directly energy consumption. And b) To raise environmental awareness of the systemic nature of changes, (affecting the sustainability of ecosystems, climate, etc) via a quest for inquiry and knowledge using data from sea water sensors. The latter has not been previously addressed due to the difficulties imposed by the nature of such IoT infrastructure that can be used for getting data in sea water.

Reinforcing the educational community on educating the new generations will create a multiplier on the overall energy reductions: promoting sustainable behaviours at school will also reflect behaviours at home. Several studies document the ability of students to influence choices made by their families related to environmental issues~\cite{schelly12}. The research interviews conducted in~\cite{powerdown} made clear that energy conservation insights learned in school can be applied at home by students and their families. Since about 27\% of EU households include at least one child under the age of 18~\cite{eurostat10}, targeted efforts of reaching families of children and young people will scale further to reach a large portion of the EU population and multiply the benefits towards sustainability of the planet.

The rest of the paper is structured as follows. In Sec.~\ref{sec:sota} relevant literature is presented and in Sec.~\ref{sec:iot} a platform for using data collected from IoT deployments at schools in STEM is presented in details. In Sec.~\ref{sec:energy} a set of educational actions is presented that uses the data collected from the IoT infrastructure to focus on promoting behaviour change. In Sec.~\ref{sec:water} a new set of educational scenarios are proposed for examining the aquatic sector and how to educate students on sustainable behaviours. The paper concludes in Sec.~\ref{sec:conlusion} where future research directions are also provided.

\section{Related work}
\label{sec:sota}
The approach of promoting sustainable behavioural change through activities targeting the educational sector falls within the scope of several research projects. \cite{smartcampus} focuses on 4 public university sites with pilot hardware installations, combined with software aimed either at desktop users or mobile users, for promoting energy consumption awareness and engagement. Focused on school buildings are the Veryschool \cite{veryschool} and Zemeds \cite{zemeds2015} projects, producing recommendation and optimization software components, or methodologies and tools. \cite{school-future} produced several guidelines and results regarding good energy saving practices in an educational setting. 

The procedures proposed (see sections \ref{sec:energy} and \ref{sec:water}) include the users monitoring data and drawing subsequent conclusions - as a part of games or school education assignments- as a first step towards raising awareness. The concept of users in the loop of monitoring is central in the area of participatory sensing \cite{Burke06participatorysensing} in which personal mobile phones of users are used to collect relevant data for a number of applications such as urban planning, public health, cultural identity and creative expression, and natural resource management. This approach has been employed by the  Cornell Laboratory of Ornithology \cite{doi:10.1080/09500690500069483} in a science education project on bird biology, while in \cite{6296848} the authors describe trials for air quality, water quality and plant disease monitoring. Similarly to our context, \cite{6990342} presents a solution combining a deployed and participatory sensing system for environmental monitoring.

Therefore facilitating the development of diverse application scenaria and supporting the different requirements in terms of data interpretation and analytics is a crucial aspect. In \cite{akrivopoulos-etfa-2017} look into people-centric applications for facilitating the educational sector towards improving the energy efficiency of school buildings. In a broader context, in \cite{DBLP:journals/iotj/GutierrezAMM18} people-centric scenaria are examined at a smart city level.

Other related approaches are reporting on IoT enabled gamification, targeting reduced energy consumption in public buildings \cite{DBLP:conf/giots/PapaioannouKBLD17} and \cite{DBLP:conf/giots/MylonasALZZHFPC17}.

An integration of an IoT data management platform and a serious game, whereby users compete in energy-related actions is reported in:
\cite{DBLP:conf/giots/Garcia-GarciaTG17}. 
In the past several approaches have been proposed in order to address the potentially huge number of sensor data arriving from the IoT domain, each one of them applied in different parts of the network architecture~\cite{Chatzigiannakis2007466,Chatzigiannakis2005376,DBLP:conf/iot/ChatzigiannakisHKKKLPRT12,DBLP:conf/anss/ChatzigiannakisKN05}. In the GAIA project approach, a platform is used for sensor reading related gamification activities, referred to as "the GAIA challenge", which can be seen in \cite{iot-newsletter}.

In \cite{Heggen:2013:PSR:2405716.2405722} the authors discuss the value of participating to project like these for students, concluding that ``Students are gaining deep domain-specific knowledge through their citizen science campaign, as well as broad general STEM knowledge through data-collection best practices, data analysis, scientific methods, and other areas specific to their project''

\section{IoT and Real-world Data in STEM}
\label{sec:iot}
One approach for addressing the climate change problem is through the development and transfer of green technologies. In the context of reducing the energy spent in residential buildings, new technologies have been introduced that improve the energy efficiency of buildings. In fact, till now the dominant approach was to use energy-efficient infrastructure and materials to reduce the energy consumption of buildings. Unfortunately, the rates of construction of new buildings as well as the rates of the renovation of existing buildings are both generally very low~\cite{bpie11} to expect a significant effect on the total amount of energy spent in our everyday life at a global level. Similarly, the approach for reducing the energy consumption in transportation focuses on improving the energy efficiency of motor engines. Also here, given the rate of change of existing fleets with energy efficient one, it is very challenging to save energy in this sector through this approach~\cite{iccs08}. 

An alternative approach, that has recently received emphasis, is the promotion of energy consumption awareness, sustainability and behavioural change on people. The main concept is that to address the global climate problem requires informing the population about their roles in mitigation actions and adaptation of sustainable behaviours. In other words, addressing climate change and achieve ambitious energy and climate targets requires a change in citizens' behaviour and consumption practices~\cite{eea13}. Reports indicate that citizens making efficient use of energy in their everyday life can lead to large energy and financial savings and potentially to a substantially positive environmental 
impact~\cite{eea13}. 

A key challenge for achieving sustainability and transforming people's behaviour towards energy consumption is the need to educate them on such issues. An interesting starting point is the educational sector. Raising awareness among young people and changing their behaviour and habits concerning energy usage is key to achieving sustained energy reductions. At EU level, people aged under 30 represent about a third of the total population \cite{eurostat12}. Thus, by targeting this group of citizens we affect a large part of the EU population. Additionally, young people are very sensitive to the protection of the environment so raising awareness among children is much easier than other groups of citizens (e.g., all attempts made to achieve behavioural change and establish new environment-friendly habits to children regarding recycling have had very high success rates).

The \gaia platform~\cite{s17102296,mylonas2018enabling,akrivopoulos2018fog,amaxilatis2017enabling} is among the very few IoT systems that have focused on the educational community. A real-world IoT deployment is spread in 3 countries (Greece, Italy, Sweden), monitoring in real-time 18 school buildings in terms of electricity consumption and indoor and outdoor environmental conditions. The data collected is used as part of series of education scenarios whose goal is to educate, influence and attempt to transform the behaviour of elementary school students through a series of trials conducted in the educational environment and in homes. Feedback mechanisms notify the students on current energy consumption at school and in this way assist towards raising awareness regarding environmental effects of energy spending and promote energy literacy by educating the users.

\gaia is based on the principle that continuously monitors the progress of students positively contributes towards reducing the energy consumption and successful behaviour change. Since the IoT deployment is multi-site and multi-country can motivate, for example, to identify energy consumption patterns in different countries and across different climate zones. This can be used to make comparisons or competitions; for instance, students of school A compete with students of school B inefficiency. This could also help understanding cultural differences with respect to energy efficiency awareness and sustainability.

The deployed devices provide 880 sensing points organized in four main categories: (1) classroom environmental comfort sensors (devices within classrooms); (2) atmospheric sensors (devices positioned outdoors); (3) weather stations (devices positioned on rooftops); and (4) power consumption meters (devices attached to the main breakout box of the buildings, measuring energy consumption). Given the diverse building characteristics and usage requirements, the deployments vary from school to school (e.g., in number of sensors, hardware manufacturer, networking technology, communication protocols for delivering sensor data, etc.). The IoT devices used are either open-design IoT nodes (based on the Arduino popular electronics prototyping platform, see~\cite{Pocero2017,akribopoulos2010web}) or off-the-shelf products acquired from IoT device manufacturers. 

The platform also incorporates participatory sensing technologies for semi-automatic periodical collection of energy usage to acquire information in buildings where no IoT sensing elements are available, e.g., utilizing web/smartphone/social networking applications for acquiring information on room occupancy, usage of conditioning or special machinery, opening of windows, etc. The goal of \gaia is to include the users in the loop of monitoring the energy consumption in the buildings they use daily, thus making the first steps towards raising awareness, connecting the educational activities carried out at schools with their activities at their home environment and also engaging the parents and relatives at home. The teacher can initiate participatory sensing sessions during the courses so that students can use phones and tablets to gather data in real time and then review them in class (for more information see Sec.~\ref{sec:energy}).

The integrated sources of input are utilized to continuously provide direct feedback, custom-tailored to each particular learner/audience (i.e., kindergarten, school, university, parents). Direct feedback is provided via real-time displays (RTDs) installed at central locations in the buildings, published on school websites, posted to social media, and also displayed on the users' smartphones and tablets. Direct feedback mechanisms are developed to address the immateriality of energy \cite{pierce2010materializing} and make it a visible entity by connecting it to the daily activities of students. Visual analytics are combined with recent advances in IoT sensing and pervasive computing technologies to provide an interactive environment that \textit{stimulates behavioural change on a frequent basis}. The energy consumption topic is included in the pedagogical activities of the schools incorporating educational aspects to promote energy literacy, convey information regarding historical data and comparative information with other buildings of similar characteristics (for more information see Sec.~\ref{sec:energy}). 

A series of social-networking applications are provided to set community-based incentives for pro-environmental behavioural change and promote collective consuming of resources. These applications utilize the already established relationship between users of the same school/department to provide \textit{community-based initiatives} to reduce their overall environmental footprint and increase environment-friendly activities. A series of \textit{game-based competitions} further engage the students in learning how to improve the energy efficiency, and to encourage them to actually follow the learned practices. Research suggests that competitions can be effective in promoting environmentally responsible behaviour \cite{gobel2010serious}. Historical data collected from the IoT infrastructure allows students to compete with each other on periodic intervals (e.g., per week/month/season) to further motivate eco-friendly behaviours. A combination of \textit{direct competition} among other groups of similar size, climate zone, socio-economic characteristics and past years (e.g., class 2016 vs recordings from class 2015) and \textit{indirect competition} against each group's own performance is followed. These competitions \textit{encourage spreading the word to larger groups}, allowing related persons, such as parents, friends, or neighbours, to participate and also \textit{appeal to positive emotions}, such as hope and enjoyment, as ways to changing individuals’ behaviours.

\myparagraph{Bringing IoT into the sea.} Most of the works in the IoT domain focus on terrestrial applications. Even when offshore infrastructures or vessels are considered, IoT devices are mostly deployed in ``dry'' surfaces and only some specific transducers are actually deployed into the water. The underwater environment is hostile, and consequently, underwater IoT devices are very expensive. If you only consider a reliable water-proof housing for shallow water, it costs at least 2 or 3 order of magnitude more than terrestrial solutions and much more if you consider deep water scenarios. Underwater operations are complex and challenging. As an example, the fast growth of algae or microorganisms can suddenly affect the quality of sensors readings that have to be often cleaned. Underwater communications are still extremely difficult and energy-hungry; RF propagates only a few centimetres and only acoustic or optical communications can be used for longer distances. The energy cost of underwater communications strongly limits the device lifetime, that is usually in the order of few months at best and requires frequent replacements of the batteries, an annoying, time-consuming and difficult task. Finally, communication standards are emerging only in the last years.
Due to these reasons, the availability of underwater IoT data is still very limited. One of the few attempts to provide a federation of underwater testbeds for the Internet of Underwater Things is the EU project SUNRISE \cite{SUNRISE}. 
While SUNRISE clearly showed us the potential of exploring underwater data, it was not originally conceived for STEM educational activities, and both the complexity of the tools and the costs of the equipment are not yet suitable to be operated by students. Despite these difficulties, there are already some efforts for more affordable tools for underwater investigations \cite{Baichtal:2015:BYO:2886330,cave} 
and is, however, possible to design significant STEM activities (see section \ref{sec:water}) that focus on shallow water and/or surface sampling that significantly lower the above-discussed difficulties. Indeed, the focus on the shallow water and/or the sea surface  allow us to a) engage students in participatory sampling (i.e. they are directly involved in the sampling procedure at sea), b) deploy relatively simple networking infrastructures capable to deliver the data acquired by possible underwater traducers employing standard wireless technologies (e.g. Lora, Sigfox or even WiFi). In the latter case, the transducers can be placed underwater and the collected data are delivered by a cable to a wireless device on the surface that makes them available in the cloud.

\section{Energy Efficiency Education}
\label{sec:energy}
A main objective of environmental sustainability education in terms of raising awareness towards energy efficiency is to make students aware that energy consumption is largely influenced by the sum of individual behaviours (at home, school, etc.) and that behaviour changes and simple interventions in the building (e.g., replacing old lamps with energy-efficient ones) can have a great impact on achieving energy savings. IoT technologies can support these initiatives by mediating people's interaction with the environment in order to provide immediate feedback and actually measures the impact of human actions while automating the implementation of energy savings policy and at the same time maintaining the comfort level perceived by people. 

Teachers can use collected data and analytics during class to explain to pupils basic phenomena related to the parameters monitored and organize student projects, where each student monitors specific environmental parameters at their home. In Monitoring school buildings situated in different countries can help, e.g., to identify usage or energy consumption patterns. This, in turn, can be utilized to make comparisons or realize competitions through social networking and game applications (e.g., students of school A compete with students of school B in answering energy awareness questions). 

Including the users in the loop of monitoring their daily energy consumption is a first step towards raising awareness. In an educational environment, this step can be further enhanced and capitalized in the framework of educational activities with the support of the IoT infrastructure. The educational activities in each school are based on data produced within the respective buildings, while the effects of changing certain behaviours can be detected and quantified. E.g., teachers can complement existing educational activities on sustainability with simple actions with immediate IoT-enabled feedback, such as turning off the lights in parts of the building and monitoring the drop in consumption or using thermal cameras to discover problematic areas combined with data showing the effect of incomplete building insulation.

\myparagraph{Scenario 1: The Importance of Building Orientation}
is the practice of facing a building so as to maximize certain aspects of its surroundings, such as street appeal, to capture a scenic view, for drainage considerations,
etc. With rising energy costs, it is becoming increasingly important for builders to orient buildings to capitalize on the Sun's free energy. In this scenario, the students are introduced to basic notions of building orientation and how to take advantage of the sun warmth to increase indoor comfort and reduce energy consumption. IoT sensors that monitor indoor temperature and humidity are used to observe how indoor conditions vary throughout the day. Data collected from the other classrooms of the school are used for comparing the indoor conditions of rooms with a different orientation. 
The educational scenario provides information on how to reconstruct the surroundings of the building in order to affect the effects of the sun. As example trees are an important factor in passive solar design because they can provide shade during hot summer days. Data collected from classrooms of similar orientation where however there are different trees located on the outside are used to observe how indoor conditions vary.

\begin{figure}[!t]
\centering
\includegraphics[width=\columnwidth]{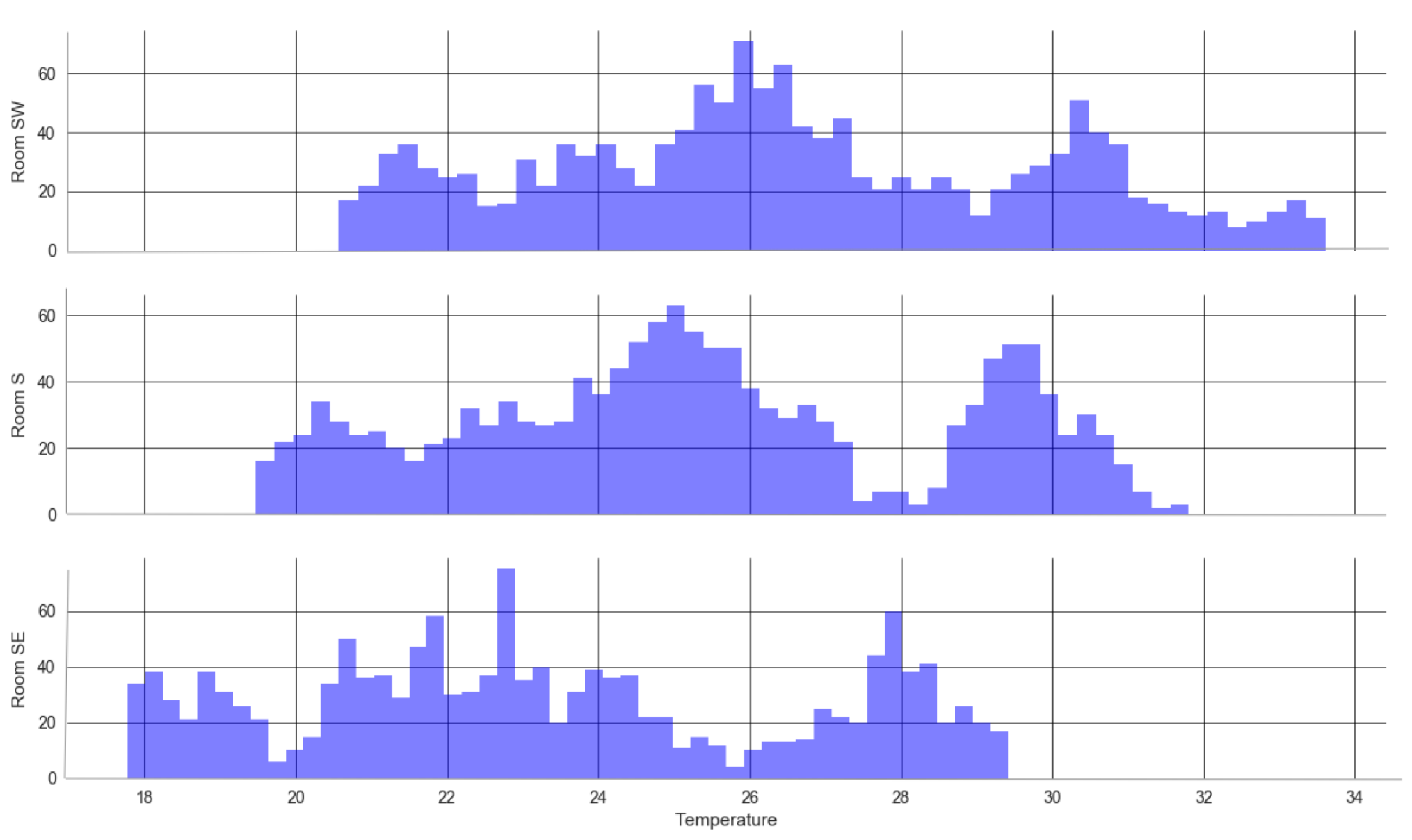}
\caption{\label{fig:histogram} Indoor temperature histogram for three classrooms during Sep/17 to Oct/17}
\end{figure}

In Fig.~\ref{fig:histogram} a histogram is provided for the indoor
temperature of the three classrooms (facing south,
south-west and south-east) examined during a period of 2 months. 
Lower temperatures are observed in the room facing South-East in contrast to the other two rooms. 
Examining the classrooms temperature is a measure of understanding the
conditions under which students and teachers operate. Hot, stuffy
rooms---and cold, drafty ones---reduce attention span and limit
productivity. 

\myparagraph{Scenario 2: Insulation Materials} have a critical impact on the indoor conditions of classrooms during the daily educational activities. Evaluating the indoor conditions of a building also requires considering other factors related to the construction materials used, the location of the windows and the heating and ventilation technology used. The \gaia platform includes school buildings located in different climatic zones, constructed in different years ranging from 1950 to 2000, using diverse materials and with different heating and ventilation systems. The education scenario uses the data collected from separate buildings in order to demonstrate the behaviour of temperature and humidity across buildings located in similar climatic zones which however have different construction methodology.

\begin{figure}
\centering
\includegraphics[width=\columnwidth]{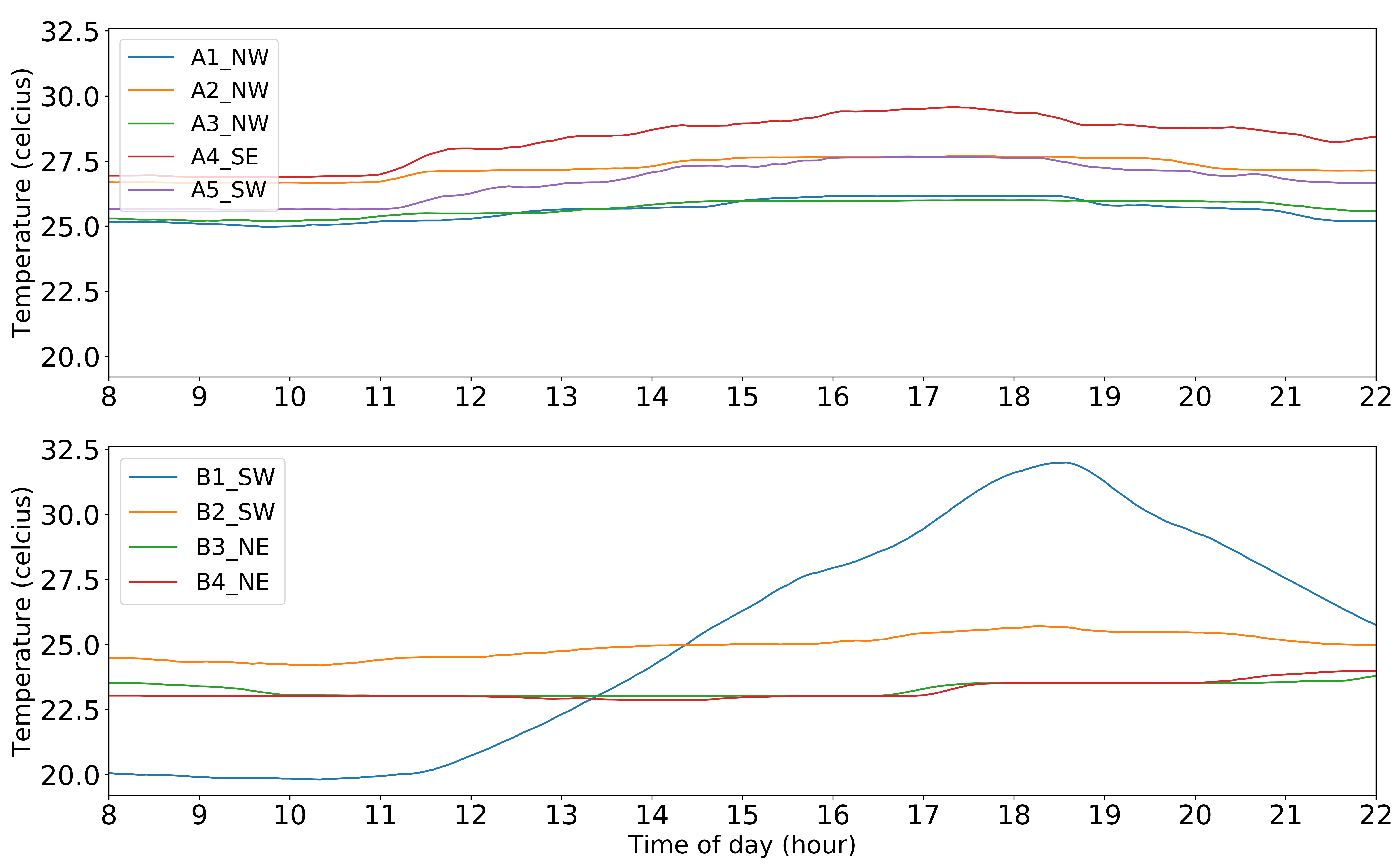}
\caption{\label{fig:weekend} Classroom temperature during 30/Sep (Saturday)}
\end{figure}

Given the above considerations, the temperature of each room is examined to identify
poorly performing classrooms. In Fig.~\ref{fig:weekend} two specific performance issues regarding two schools located in the same city are depicted. The first issue is related to the bottom figure, 
where room R1 achieves very poor performance with temperature starting at the very low level of $20^oC$ 
and increasing up to $32^oC$ within 8 hours. The second issue is related with the top figure, 
where the south-west facing classroom (R5) and the south-east facing classroom (R4) have an increase of 2 degrees during the day while all the other rooms are not affected. Even the south-west
facing room R2 of the bottom figure does not have such an increase during the day. 
After contacting the school building managers it was reported that (a) room R1 (bottom school)
is located outside the main building, within a prefabricated iso box where insulation is very poor and
(b) rooms of top school have no window blinds installed in contrast to the bottom school where window blinds are installed in all rooms. These are just two examples of the results of the analysis conducted. It is expected that such an analysis can provide strong evidence on how to improve the performance of schools.

During spring 2017, a set of preliminary testing was conducted over several weeks to get feedback regarding the educational scenaria that promote energy efficiency and sustainability. A total of $944$ students and teachers had the first interaction with the \gaia platform, while we conducted a form-based survey focusing on the gamification component ($196$ high-school students in Sweden and Italy) and the Educational Lab Kit ($132$ 6th graders in Greece). With respect to the game, 78\% of the students found the content interesting (21\% “extremely”, 26\% “very”, 31\% “moderately”) and 89\% the activity user-friendly (38\% “extremely”, 29\% “very”, 22\% “moderately”). Regarding the acceptance of the tools from educators, the direct response gathered through workshops has been positive and several schools have provided their own schedules for integrating \gaia tools in classes. Thus, in terms of overall acceptance of both the tools and the infrastructure inside buildings and the schools’ curricula, the results indicate that the educational scenaria had a quite positive response.

\section{Sea Polution Education}
\label{sec:water}
Most of the planet's surface is covered by the sea. Specifically, about 79\% of the surface of the Earth is covered by water and only 21\% of the land. Today, we know the great importance of the sea for life on the entire planet and especially for humans. The sea has a multiple importance as being a ``source of life" for Earth. It provides the ability to produce food, minerals and energy, is a key factor for the renewal of the oxygen we breathe, and the means of transporting goods (trade, energy transfer/information). Maritime trade routes have also been cultural bridges, integrating culturally large and disperse geographic areas and allowing the development of cultures. However, in order to achieve better use of the potential of the sea and at the same time to effectively protect it, a detailed study is required~\cite{sea12}.

In the educational scenarios proposed, a series of sensors are used to measure physical and chemical marine parameters. As already observed in section \ref{sec:iot}, bringing the IoT into the sea is still very difficult, for this reason, we will focus on surface sampling activities that are more affordable in the context of STEM educational activities.

The steps of the pedagogical activities we follow are awareness, observation, experimentation and action. School students located in Europe's coastal areas use portable equipment to carry out relevant measurements and submit them to a database they have access to.
Depending on the teaching needs and priorities, students can collect and analyze the following:

\begin{itemize}

\item current  values and any fluctuations  of them during the observation period of the activity,
    
\item changing  values for longer periods of time, e.g. making comparisons between different times of the day, between months, seasons, or years,

\item the variance of the phenomena between different areas.
    
\end{itemize}    
    
The mathematical and scientific thinking developed in the above process can be exploited in various ways by the tutor, in the context of teaching mathematical and scientific skills,  not only in the science courses but also in cross-thematic approaches that combining such observations and analyzes the economic, social and other aspects of our effort for clean seas.

\myparagraph{Scenario 1: Observation of Sea Water Temperature} is achieved via water temperature sensors positioned at the surface of sea level. Surface water temperature has a natural daily (diurnal) and seasonal variation due to weather conditions and thermal exchanges with the atmosphere. Students use the IoT infrastructure during the school year to observe the temperature of the surface water of the sea, examining measurements at different times of the day and at different depths (up to $10m$) and experimentally confirm their theoretical predictions.

Given that \gaia platform is deployed across different countries, \textit{the data collected can accommodate the study of the surface temperature of the sea in relation to latitude}. The surface temperature distribution fully corresponds to the distribution of the solar radiation entering the sea. The global ocean surface temperatures (for water depths up to $5m$) show a bandwidth in terms of latitude. Near the Equator, the waters have high temperatures throughout the year. On the contrary, in the areas near the poles, the temperatures of the surface layers are almost always very low. In the temperate climate zone, the temperature values obtained by the surface water mass, are lower than those of tropical waters and higher than the corresponding polar waters and change significantly during the year~\cite{iccs08}.

\begin{figure}
\centering
\includegraphics[width=\columnwidth]{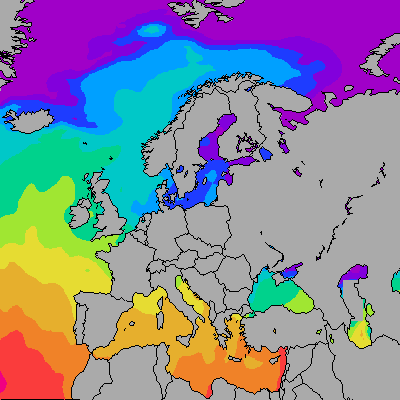}
\caption{\label{fig:sea-temperature} Surface temperature of the seas of Europe}
\end{figure}

Students in the Mediterranean coastal regions will record the higher surface temperature on the same day of the year than pupils in the Baltic or Atlantic coastal regions, due to the different latitude and hence to different amounts of solar radiation that the region receives.

\textit{The historic records collected from the IoT infrastructure also enable to study the temperature of the sea-surface during the course of a year.} Students using the measurements they recorded during the year will be able to explain the seasonal variation of surface temperature. In winter, the waves of the sea are more intense and the surface layer is being mixed while the temperature is low and uniform. In the summer, where the atmospheric temperature is high and the wave intensity is small, water mixing is minimal and the temperature of the surface layer increases strongly due to heat build-up. From March to August, the temperature on the surface of the sea is constantly increasing due to the absorption of heat from the atmosphere. So, in the spring, the sea surface begins to create seasonal thermocline, which has a small thickness and small temperature range. In summer, the surface temperature is constantly increasing due to the high temperatures prevailing in the atmosphere, resulting in the seasonal thermocline becoming thicker and higher in the temperature range. From September to February, the surface layer is constantly losing heat. The temperature of the atmosphere is smaller than the sea, and the intensity of the waves is constantly increasing so that it is fully agitated. So in autumn the thermocline thickness increases, but its width decreases relative to summer. In winter, the decrease of the surface layer continues, with the result that the temperature becomes uniform up to the ceiling of the permanent thermocline.

The annual range of surface temperatures is maximal at the intermediate latitudes, while at the small and large geographical heights the range is minimal. School pupils in medium latitudes, due to the high fluctuations in atmospheric temperature and other atmospheric phenomena, expect to record a higher time variation in the temperature of the surface layer. 

The educational scenario can be extended to include measurements in shallow coastal areas, where seasonal variation may be absent from the general rule, particularly if there is a significant effect of brackish waters. The students conclude that local water mass inflows and the resulting mixing of different water types, a phenomenon particularly common in coastal waters, may lead to deviations from the general rules.

Finally, the capability of the IoT infrastructure to \textit{continuously collect data in real-time allows the study of the surface temperature of the sea during the day.} Apart from the seasonal thermocline in the middle latitudes, there is also the daily thermocline (diurnal), which is particularly pronounced in the spring, summer and autumn. Groups of students from schools in different coastal regions of Europe collect daylight surface temperature data, record the values ​​in the database, process the data, and arrive at scientific conclusions about the variation in surface sea temperature.

\myparagraph{Scenario 2: Observation of Sea Water Acidity-Alkalinity} is achieved by deploying \textsf{pH} sensors\footnote{Gravity: Analog pH Sensor / Meter Kit For Arduino} across different coastal sites. \textsf{pH} plays a major role in the marine ecosystem because it determines the solubility and chemical form of most of the substances present in it. The reduction or increase in \textsf{pH} is directly related to the photosynthesis and respiration of the marine ecosystem organisms and therefore is related to the productivity of the biomass. On the surface of the sea the \textsf{pH} ranges from $8.0$ to $8.3$ and depends on the atmospheric pressure of the $CO$, the temperature and the salinity of the water. Students find that the chemical properties of seawater differ from those of the sweet because of the presence of salts. The less acidic salts contained in seawater (such as carbonates, bicarbonates and borates) reduce the high acid or alkaline composition of any liquid waste. So the toxicity of the wastewater is high in the freshwater and decreases in the sea. \textsf{pH} measurement is the best way to assess the effects of acid or alkaline waste disposal on the marine ecosystem.

An additional goal of this activity is for students to understand that the critical survival limit for life in lakes and water streams does not depend on the average value of \textsf{pH} (degree of accelerating) over a year but on the lowest value of \textsf{pH}. Such short but dangerous periods with low \textsf{pH} values occur, mainly in the spring during the melting of the ice (acidity shocks). Fluctuations in \textsf{pH} may result in the death of many organisms (e.g., plankton at $6.5$ and perch and eel at $6.4$ and $6.3 - 6.5$ respectively). If the \textsf{pH} value is below $6.5$, the adverse effects on all living organisms begin and below \textsf{pH} $5$ all animals and plants die.

\myparagraph{Scenario 3: Observation of Sea Water Salinity and Conductivity} is achieved by deploying salinity sensors\footnote{Vernier Salinity Sensor} (\textsf{S}) and conductivity sensors\footnote{Vernier Conductivity Probe} (\textsf{STD}) across different sites of the IoT deployment. Students in different coastal regions of Europe record sea salinity values and conclude that the total mass of dissolved salts varies from one sea to another, exceeding $36 grams$ in the Mediterranean Sea and falling below $10$ grams of salt per kg of water in some areas of the Baltic Sea. Students conclude that surface salinity is greatest at latitudes where annual evaporation is greater than annual rainfall and minimum salinity values are found at latitudes where rainfall is greater than evaporation.  River water also affects salinity values, for example, the Baltic Sea is a basin with limited communication with the Atlantic, where large rivers are poured, while evaporation is minimal. On the contrary, the Mediterranean and the Red Sea are two basins where the exhaust is large and the discharge of river water is minimal, resulting in large amounts of salinity. In addition, melting and ice formation plays a role in Polar Regions.

\myparagraph{Scenario 4: Observation of Sea Water Turbidity} is achieved by deploying turbidity sensor\footnote{Vernier Turbidity Sensor} (\textsf{T}) across different sites of the IoT deployment. Students using the turbidity sensor record physical parameter values ​​that determine the ability of sunlight to pass through the water. Turbidity is caused either by natural causes (erosion or decomposition of organisms after their death) or by the colloidal and fine-grained suspended solids contained in sewage and industrial waste and precipitating at the bottom with great difficulty and directly affecting ecosystem species with increased need light for their development. The depth of penetration of light in seawater is critical for primary production (photosynthesis) and depends on the clarity of seawater and the wave of light radiation.

\myparagraph{Scenario 5: Observation of Dissolved Oxygen Sensor} is achieved by deploying dissolved oxygen sensors\footnote{Vernier Dissolved Oxygen Probe} (\textsf{DO}) across different sites of the IoT deployment. 
Wastewater from our houses contains organic substances that can be used as feed by other organisms, particularly microbes. These organisms with oxidative reactions metabolize organic substances by consuming for this process the oxygen dissolved in the water. Because oxygen has relatively little water solubility, it is quickly consumed when there is a high organic load resulting in anaerobic conditions. Concentration less than $7 mg/lt$ means oxygen deficiency resulting in the non-survival of fish and other aerobic organisms. Physiological values of \textsf{DO} range above $7 mg/lt$. Sensors of turbidity and dissolved oxygen will be used to study the phenomenon of eutrophication on closed shores - gulfs where water circulation is limited, near coastal rural areas or in areas near ports or in areas where sewage flows into the marine environment without being biologically cleaned. Apart from the areas where eutrophication affects the environment, students will also identify areas such as river estuaries, which tend to be naturally eutrophic, because they transport nutrients to the open sea, giving increased productivity and food to fish and other organisms. It is not by chance that important sea fishing grounds are located near estuary areas (North Aegean, Thracian Sea, etc.).

\section{Conclusions}
\label{sec:conlusion}
This paper reports on the use of  IoT sensors as a basis for educating children in environmental awareness and STEM. A number of educational scenarios are presented based on sensors readings from school buildings as well as seawater measurements. The paper raises the subject of early awareness of environmental issues based on real-world data and the use of motivating / gamified scenarios.

An education-focused real-world IoT deployment in schools in Europe can help promote sustainable activities~\cite{1612844}. By using this infrastructure and the data it produces, it is easier and more effective to build tools that better reflect the everyday reality in school buildings and provide a more meaningful feedback. The development of educational activities focusing on energy awareness in schools has received very positive feedback from the educational community. We believe that recent technological developments allow us to extend the IoT infrastructure in order to monitor additional environmental parameters apart from energy consumption~\cite{TZIORTZIOTI2019117,ami2018}. Towards this end, we propose the deployment of sensors in the sea to form underwater sensor networks for monitoring the aquatic sectors of our planet. A series of educational scenaria that utilize the collected data to further promote sustainability awareness and behavioural change. This can be achieved via educational activities in schools as well as gamification, facilitated by platforms such as the one realized in the GAIA project.

\section{Acknowledgments}

This work has been partially supported by the EU research project ``Green Awareness In Action'' (GAIA), funded under contract number 696029 and the research project Designing Human-Agent Collectives for Sustainable Future Societies (C26A15TXCF) of Sapienza University of Rome.
This document reflects only the authors' view and the EC and EASME are not responsible for any use that may be made of the information it contains. 

A previous version of this paper has appeared in the \textit{2018 IEEE Conference on Computational Intelligence and Games (CIG)}, DOI: \url{https://doi.org/10.1109/CIG.2018.8490370}, \cite{8490370}. \textcopyright 2018 IEEE.  Personal use of this material is permitted.  Permission from IEEE must be obtained for all other uses, in any current or future media, including reprinting/republishing this material for advertising or promotional purposes, creating new collective works, for resale or redistribution to servers or lists, or reuse of any copyrighted component of this work in other works. 


\end{document}